\documentstyle[12pt]{article}
\begin{document}

\def\p{\phi}
\def\P{\Phi}
\def\a{\alpha}
\def\e{\varepsilon}
\def\be{\begin{equation}}
\def\ee{\end{equation}}
\def\l{\label}
\def\0{\setcounter{equation}{0}}
\def\T{\hat{T}_}
\def\b{\beta}
\def\S{\Sigma}
\def\3{d^3{\rm \bf x}}
\def\4{d^4}
\def\C{\cite}
\def\r{\ref}
\def\ba{\begin{eqnarray}}
\def\ea{\end{eqnarray}}
\def\n{\nonumber}
\def\R{\right}
\def\L{\left}
\def\q{\hat{Q}_0}
\def\X{\Xi}
\def\x{\xi}
\def\la{\lambda}
\def\d{\delta}
\def\s{\sigma}
\def\f{\frac}
\def\vx{{\rm \bf x}}
\def\j{\frac{\delta}{i \delta j_a ({\rm \bf x},x_0+t+t_1)}}

\begin{titlepage}
\begin{flushright}
{\normalsize IP GAS-HE-9/95}
\end{flushright}
\vskip 3cm
\begin{center}
{\Large\bf The path integrals on a curved manifold}
\vskip 1cm

\mbox{J.Manjavidze}
\footnote{Institute of Physics,
Georgian Academy of Sciences, Tamarashvili str. 6,
Tbilisi 380077, Republic of Georgia,
e-mail:~jm@physics.iberiapac.ge} \\

\end{center}
\date{april 1995}
\vskip 1.5cm

\begin{abstract}
\footnotesize

For description of the quantum dynamics on a curved group manifold the
path integrals in a space of the group parameters is offered. The
formalism is illustrated by the $H$-atom problem.

\end{abstract}
\end{titlepage}

\section{Introduction}
\setcounter{equation}{0}

The classical particles motion may be described mapping the trajectory
\footnote{We shall start the consideration from the flat space} on the Lee
group $G$ manifold \C{cart} imagining the particles coordinates as the
elements of $G$. Then, roughly speaking (e.g. \C{fin}), the group
combination law creates the particles classical trajectory. On the
homogeneous and isotropic group manifold the motion of a particle is free,
it mooves with the constant velocity in an arbitrary point of the manifold.
It is the crusial point of this approach since actually solves a quantum
problem.

Indeed, this idea was used for description of the particles quantum motion
\C{dow} and it was shown for the essentially nonlinear Lagrangian
\be
L=\frac{1}{2}g_{\mu \nu}(x)\dot{x}^{\mu}\dot{x}^{\nu}
\l{1.1}
\ee
that the quasiclassical approximation is exact on the (semi)simple Lee
group manifold.

But this slender mechanism of solutin is destructed in presence of
interaction potential $v(x)=O(x^n)$, $n>2$, since it breaks the isotropy and
homogeneity of the Lee groups manifolds \C{cart}.

We offer another approach to include the interaction potential. Firstly, we
shall specify the topology of group manifold $M_s$ through the group of
the corresponding classical equation of motion. So, we shall assume  that
this problem has the classical solution.

But in the quantum theory one must expect that the quantum perturbations
should lead to deformations of $M_s$. This is the main problem: one must
quantize the manifold $M_s$ to describe the motion in the unhomogeneous
and nonisotropic space.

Instead of this complicated problem we wish count the (quantum) fluctuations
of the parameters of group manifold (GMP). It had shown in the paper \C{un}
that this is possible (Sec.2) since conserves the integral probability.

It is remarkable that this ``unitary" definition of the functional
measure allows to consider independently the quantum exitations of each
independent degree of the freedom of the invariant manifold $M_s$, i.e. of
each independent parameters of the classical trajectory.

In result, the GMP will be defined through the generators of $G_s$.
Considering the parameters of invariant hypersurface as the generalized
coordinates and momenta of the ``particles", in the classical limit the
motion of this ``particles" must be free \C{arn}, i.e. theirs velocities
are constants. Such choice of the ``particles" coordinates allows to
achive the same effect as in the above discussed transformation to the
homogeneous and isotropic (semi)simple Lee group manifold \C{cart}.
Moreover, even in quantum case one can get to the free ``particles" motion
rescaling the quantum sources.

Considering the quantum problem on the compact manifold one can demonstrate
the absence of quantum corrections to the angular degrees of fredom. This is
the new solution of the quantum integrability problem \C{ang}. This result
seems important since it demonstrates also the possibility of the partial
quasiclassicality (i.e. of the partial integrebility) of a quantum problem.

It is known that the dynamical groups are more rich \C{ibr}
and they cannot be reduced to the Lee groups. This forces us to consider
the quantum problems in a phase space since without this groups one
cannot solve a number of modern field-theory problems.

It is evident that the mapping on the dynamical group manifold has not a
simple geometrical interpretation as we had discussed above for the Lee
groups (the trajectory can not belong to the manifold of the dynamical
group). But formaly, as it will be seen, the description of quantum motion
on the ``dynamical groups manifolds" is practically the same: we will deal
with the space of parametres of the dynamical group.

Therefore, the offered method is applicable as for the Lee groups,
so for the dynamical groups since in both case we will work in the space of
group parameters. It is better to say that we will consider the
classical trajectories manifold. This allows to extend the aproach on the
field-theoretical problems.

To do the method clear we shall consider the $H$-atom problem. This problem
has the additional (hidden) first integral in involution, the
Laplace-Runge-Lentz (LRL) vector:
\be
\vec{A}=(\vec{l}\times \vec{p})+\vec{x}/|x|,
\l{1.2}
\ee
where $\vec{p}$ is the momentum and $\vec{l}$ is the angular momentum.
It is easy seen that pare ($\vec{A},~\vec{l}$)  form the algebra of
$O(4)$ Lee group. It allows to solve the $H$-atom problem algebraically
\C{pauli}. Transforming the Schrodinger equation to the momentum space one
can see the $O(4)$ symmetry of this equation. This allows to
solve the transformed equation and to explain the degeneracy of the energy
terms \C{fock}.

The isomorfism between $H$-atom problem and harmonic oscillator model was
shown in \C{duru} using the path-integral formalism. One can interpret this
result as the absence of necessity to handl the $O(4)$ symmetry for the
$H$-atom problems integrability.

For us the crucial point consist
in the fact that the particles trajectories in the Coulomb potential are
closed independently from the initial conditions. It is known that this
is the result of conservation of the LRL vector \C{bert}.

We map the $H$-atom problem initially defined in the plain Cartesian
space into the space of the group parameters (Sec.2). The trajectory can be
defined for $O(4)$ group by the pare ($A,~l$) and by the true anomaly
$\phi_t$, or for $O(3)$ group by the pare ($h,~l$), where $h$ is the
classical Hamiltonian, and by the eccentric anomaly $\phi_e$. We will choose
the second definition and will describe the quantum fluctuations of
($h,~l,~\phi_e$).

We will show the integrability of $H$-atom proroblem demonstrating the
sufficience of angular corrections cancelation. For this we will use the
trajectory  closeness only, and presence of the LRL vector will not be used
explicitly, i.e. the presence of $O(4)$ dynamical group will be taken into
account latently.

\section{The dynamics in a curved space}
\setcounter{equation}{0}

We shall start from the flat Cartesian coordinates. The amplitude with enrgy
$E$ has the form:
\be
A(x_1 ,x_2 ;E)=i\int^{\infty}_{0} dT_+ e^{iET_+}
\int_{x_+(0)=x_1}^{x_+(T_+)=x_2} Dx_+ e^{iS_{C_+ (T_+)}(x_+)},
\l{1}
\ee
where $x$ is the 2-dimensional vector. The action
\be
S_{C_+ (T)}(x)=
\int_{C_+ (T)}  dt (\frac{1}{2}\dot{\vec{x}}^2
+\frac{1}{|x|})
\l{}
\ee
and the measure are defined on the Mills' contour \C{mil}:
\be
C_{\pm} (T): t\rightarrow t\pm i\epsilon,
{}~~~\epsilon\rightarrow +0,
{}~~~0\leq t\leq T.
\l{}
\ee
Indrocution of Mills' time contour is necessary to ensure the convergence
of path integral (\ref{1}).

We will calculatethe the probability
\be
R(E)= \int dx_1 dx_2 |A(x_1 ,x_2 )|^2 ,
\l{2}
\ee
to introduce the unitary definition of path-integral measure
\C{manj}. Insering (\ref{1}) into (\ref{2}) we find that
\be
R(E)=
\int^{\infty}_{0} dT_+ dT_-  e^{iE(T_+ -T_- )}
\int_{x_+(0)=x_-(0)}^{x_+(T_+)=x_-(T_-)} Dx e^{iS_{C^{-}} (x)}
\l{3}
\ee
is described by the closed-path integral.

The  total action
\be
S_{C^{-}} (x)=
S_{C_+ (T_+ )} (x_+ )
- S_{C_{-}(T_- )} (x_- )
\l{}
\ee
containes the sum of two actions, where $S_{C_+ (T_+ )} (x_+ )$ describes
forward in time motion and $-S_{C_{-}(T_- )} (x_- )$ backward one.
Note that $S_{C_{-}}$ is defined on the complex conjugate contour
$C_-^* (T)=C_+ (T)$ and, since $x_+(t)$ and $x_-(t)$ are in our formalism
the independent trajectories, $S^{-}$ is the complex quantity.
This allows to define the measure $Dx$: extracting the linear over
$e(t) = (x_+ -x_-)(t)/2$ term in the exponent we explicitly find
in result of integrationm over $e(t)$ that the path integral differential
measure over $x(t)=(x_+ +x_-)(t)/2$ is $\d$-like.

The path integral of (\ref{3}) type was considered in \C{un}.
Performing the transformation to the cylindrical coordinates we find that
the trajectory winds the surface arbitrary times since we map the
noncompact space on the compact one. One can choose the principal domain
for the angular variables and calculate sum over repeated contributions.

So, in the cylindrical coordinates the result looks as follows:
\ba
R(E)=2\pi \int^{\infty}_{0}dT
\exp\{-i\int_{C(T)}dt (\hat{j}_{r}(t)\hat{e}_{r}(t)+
\hat{j}_{\phi}(t)\hat{e}_{\phi}(t))\}\times
\n \\ \times
\int D^{(2)}M(r,\phi) \exp\{-iV_T (x,e_{C})\},
\l{4}
\ea
where
\be
\int_C =\int_{C_+}+\int_{C_-}
\l{}
\ee
and the Dirac's $\delta$-like measure
\ba
D^{(2)}M(r, \phi ) =\delta (E -H_T (r,\phi ))
\prod_t r^2 (t)dr(t)d\phi (t)\times
\n \\ \times
 \delta (\ddot{r}-\dot{\phi}^2 r+v'(r)-j_{r})
\delta(\partial_t (\dot{\phi}r^2 )-j_{\phi})
\l{}
\ea
($H_T$ is the classical Hamiltonian at time moment $T$). Note that
we are not able to shift the time contours $C_{\pm}$ on the real axis
since of the Green functions singularities.

The symbol ``hat" in (\ref{4}) meance the differentiation over corresponding
quantity:
\be
\exp\{-i\int_{C(T)}dt (\hat{j}_{r}(t)\hat{e}_{r}(t)+
\hat{j}_{\phi}(t)\hat{e}_{\phi}(t))\}
\equiv \exp\{-i\int_{C(T)}dt (
\frac{\d}{\d j_{r}(t)}\frac{\d}{\d e_{r}(t)}+
\frac{\d}{\d j_{\phi}(t)}\frac{\d}{\d e_{\phi}(t)})\}
\l{5}
\ee
Here $j_r$ and $j_{\phi}$ are the radial and angular perturbation sources,
and $e_r$ and $e_{\phi}$ are the corresponding auxiliary fields. At the
end of calculations one must take this quantities equal to zero. This
means that the quantum exitation sources are switched on adiabaticaly and
the equations of motion can be solved perturbatively, expanding over $j_r$
and $j_{\phi}$.

The action of the perturbations generating operator (\ref{5}) on the weight
functional
\ba
-V_T (x;e_c )=S_{C_+ (T)} (x+e_c )-S_{C_- (T)}(x -e_c) +
\n \\
+\int_{C(T)} dt (\ddot{\vec{x}} -\vec{x}/|\vec{x}|^3 )e_c
\l{6}
\ea
generates the asymptotic series \footnote{Note that only the potential
$v(x)=O(x^s),~~ s \leq 2,$ the Dirac's measure is free from the source
of quantum exitations $j$. In this case the quasiclassical approximation is
exact. Starting from Cartesian flat coordinates frame we can map this model
on the arbitrary manyfold. In result, considering the Fourier-transform of
the functional $\d$-function, we will find explicitely the model
Lagrangian (\ref{1.1}). The more complicate case will be considered.}.

The vector $\vec{e}_c$ has the components:
\be
e_{c,1} =e_r \cos\phi -re_{\phi} \sin\phi,~~~~~
e_{c,2}=e_r \sin\phi +re_{\phi} \cos\phi.
\l{}
\ee
and it was assumed that $V_T (x,e_c )$ in (\ref{6}) was written in the
cylindrical coordinates.

As was explained in the Sec.1 we will deal with motion in the tangent space.
Inserting
\be
1=\int Dp Dl \prod_{t}
\delta  (p-\dot{r}) \delta (l-\dot{\phi}r^2)
\l{}
\ee
into (\ref{4}) we introduce the motion in the phase space with
the Hamiltonian
\be
H_{j}=
\frac{1}{2}p^2 +\frac{l^2}{2r^2}-\frac{1}{r}-j_{r}r-j_{\phi}\phi.
\l{}
\ee
The  Dirac's  measure becomes four dimensional:
\ba
D^{(4)}M(r, \phi ,p,l) =\delta (E -H_T (r,p,l))
\prod_t dr(t)d\phi (t) dp(t) dl(t)\times
\n \\ \times
\delta (\dot{r}-\frac{\partial H_j}{\partial p})
\delta (\dot{\phi}-\frac{\partial H_j}{\partial l})
\delta (\dot{p}+\frac{\partial H_j}{\partial r})
\delta (\dot{l}+\frac{\partial H_j}{\partial \phi}).
\l{}
\ea
Rescaling the auxiliary field $e_r$ we take away the $r^2$ term from
the measure.

Now we will map the problem into group parameters space. Instead of
$(p,~l,~r,~\phi)$ we will introduce the motion into
$(h~,l,~\theta ,~\phi)$ space where
\be
h=\frac{1}{2}p^2 +\frac{l^2}{2r^2}-\frac{1}{r}
\l{7}
\ee
is the classical Hamiltonian and
\be
\theta =\int^r dr \{2h-\frac{l^2}{2r^2}+\frac{1}{r}\}^{-1/2}
\l{8}
\ee
is the ``time" variable. This transformation is not canonical and
\ba
D^{(4)}M(h~,l,~\theta ,~\phi)=
\d (E-h(T))
\prod_t dh(t) dl(t) d\theta (t) \phi (t)\times
\n \\\times
\d (\dot{h}-\frac{l}{r_c^2}j_{\phi}-p_c (r_c)j_r )
\d (\dot{\theta} -1+\frac{\partial r_c}{\partial h}j_r +
\frac{1}{p_c (r_c)}\{ \frac{l}{r_c^2}\frac{\partial r_c}{\partial h}+
\frac{\partial r_c}{\partial l}\} j_{\phi})\times
\n \\ \times
\d (\dot{\phi}-\frac{l}{r_c^2})
\d (\dot{l}-j_{\phi}),
\l{}
\ea
where $r_c =r_c (h,l;\theta)$ is the solution of eq.(\ref{8}) and
$p_c (r)$ is the solution of eq.(\ref{7}).

Shifting now the time contours $C_{\pm}$ on the real axis \C{un} we find:
\ba
R(E)=2\pi \int^{\infty}_{0}dT
\exp\{-i\int_{C(T)}dt (\hat{j}_{r}(t)\hat{e}_{r}(t)+
\hat{j}_{\phi}(t)\hat{e}_{\phi}(t))\}\times
\n \\ \times
\int D^{(4)}M(h~,l,~\theta ,~\phi) \exp\{-iV_T (r_c,e_{C})\},
\l{9}
\ea
where
\ba
V_T (r_c,e_{C})=S_o (r_c )+ \int^{T}_{0} dt [
((r_c +e_r )^2 +r_c^2 e_{\phi}^2 )^{-1/2}-
\n \\
-((r_c -e_r )^2 +r_c^2 e_{\phi}^2 )^{-1/2}+
2e_r r_c^{-2}]
\l{10}
\ea
and $S_o (r_c )$ defines the nonintegrable phase factor \C{manj}. The
quantization of this factor determines the bound state enrgy (see below).
Such factor will appear in the case if the phase of amplitude can not
be fixed (as, for instance, in the Aharonov-Bohm case).

If we rescale the auxiliary field $e_c \rightarrow r_c e_c$ and introduce
a new time variable $dt \rightarrow r_c dt$ the $H$-atom problem is solved
in the quasiclassical approximation exactly since it becomes isomorfic to
the harmonic oscillators case. This solution was firstly demonstrated
in \C{duru}.

But we wish to demonstrate another solution. For this purpose
we need only one fact: the periodicity of $V_T$. Firstly, taking into account
(\ref{10}) we can put out the engular quantum fluctuations over $\phi$.
This means that
\be
e_{\phi}=0, ~~~~~j_{\phi}=0.
\l{}
\ee
In result,
\be
V_T (r_c,e_{C})=S_o (r_c )+ \int^{T}_{0} dt [
\frac{1}{r_c +e_r }-\frac{1}{r_c -e_r}+
2e_r r_c^{-2}]
\l{11}
\ee
and
\ba
D^{(4)}M(h~,l,~\theta ,~\phi)=
\d (E-h(T))
\prod_t dh(t) dl(t) d\theta (t) \phi (t)
\d (\dot{h}-p_c (r_c)j_r )\times
\n \\ \times
\d (\dot{\theta} -1+\frac{\partial r_c}{\partial h}j_r )
\d (\dot{\phi}-\frac{l}{r_c^2})
\d (\dot{l}),
\l{}
\ea
Note that now $l$ and $\phi$ are classical quantitiues and $h, \theta $ are
quantum.

We can use now the closeness of classical trajectory $r_c$.
Deviding the integral over $\theta$ on the periods ($\theta$ is
proportional to eccentic anomaly) we easely
find the canselation of quantum corrections to $\theta$:
\be
j_r =0,~~~~~e_r =0.
\l{13}
\ee
This gives the quasiclassical solutin of the $H$-atom problem:
\be
R(E)=2\pi \int^{\infty}_{0}dT
\int D^{(4)}M(h~,l,~\theta ,~\phi) e^{-iS_o (r_c)},
\l{14}
\ee
where
\be
D^{(4)}M(h~,l,~\theta ,~\phi)=
\d (E-h(T))
\prod_t dh(t) dl(t) d\theta (t) \phi (t)
\d (\dot{h})
\d (\dot{\theta} -1)
\d (\dot{\phi}-\frac{l}{r_c^2})
\d (\dot{l}),
\l{15}
\ee

In result, using the identity \C{manj}:
\be
\sum^{+\infty}_{-\infty} e^{inS_1 (E)} =
2\pi \sum^{+\infty}_{-\infty}\d (S_1 (E) - 2\pi n),
\l{16}
\ee
where $S_1 (E)$ is the action over one classical period $T_1$
($S_o =nS_1$):
\be
\frac{\partial S_1 (E)}{\partial E}=T_1 (E),
\l{}
\ee
we find from (\ref{14}) the valid exression:
\be
R(E)=\pi \Omega \sum_{n} \d (E + 1/2n^2)
\l{}
\ee
where $\Omega$ is the phase space volume:
\be
\Omega =\int dl_0 d\phi_0 d\theta_0 .
\l{}
\ee
This infinte coefficient should be canceled by the normalization condition
of $R(E)$.

\section{Concluding remarks}
\setcounter{equation}{0}

In this and in the previouse paper \C{ang} there was given the attempt to
demonstrate the dynamical mechanism of depression of quantum corrections
to the quasiclassical approximation. It is noticeable that the mechanism
of depression is due to the compactness of the classical trajectories
manifold

The idea based on the observation
that known integrable quantum-mechanical problems can be solved
quasiclassically \C{un}. This mechanism seems natural remembering also the
stochastic nature of quantum trajectories.

Note also that the ``level" of integrability of the problem shifts
singularities in the interaction constant complex plane. Indeed, for the
nonintegrable case the singularities
are  located at the origin \C{bw}. This picture shows the close connection
with the threshold singularities of the amplitude \C{t'hooft}. In the
semi-integrable case the singularities at origin are canceld \C{ush} and
the main (rightest) singularities are located at the finite negative values
of interaction constant. And, at the end, the singularities of the
integrable systems are located infinitly far from the origin.

This qualitative conclusions confirms the main result of this paper
anf of \C{ang} that the integrability of quantum system meance the
stability of the classical trajectory manifolds against quantum exitations.

It is known that the case of totaly integrable systems is very rear in the
Nature. But our secondary result is the observation that a quantum
problem can be integrable over part of the degrees of freedom. This fact
should have the important consequences in the field theory.

The partial integrability of the quantum system means the stability
of the manifold over some
part of the classical degrees of freedom. In quantum field
theory absence of quantum corrections leads to absence of radiation
\footnote {This statement simply follows from the $S$-matrix unitarity
condition, or can be shown by the explicit calculations using the
generalization of LSZ reduction formula \C{s-int}. It will be
demonstrated later also.}
and the partial integrability means the impotence of the system
concerning radiation of the corresponding degrees of freedom.

Two opposite experimental fact, the confinment of the color charge and
the observation of color jets, forced to think that the Yang-Mills theory
is partly integrable. The proof of stability of the non-Abelian centre
of group manifold can solve this problem. It ensure impossibility to
observe the free color charge leaving main part of the theory quantum.

\vspace{0.2in}
{\Large \bf Acknowledgement}
\vspace{0.2in}

I would  like to thank my colleagues in the Institute of Physics (Tbilisi),
and especially A.Ushveridze, for interesting discussions and I.H.Duru for
the interesting comments. I would like also
to thank M.Tomak for kinde hospitality in the Physics Department at
METU, where the paper was finished.
This work was supported in part by the U.S. National Science Foundation.

\end{document}